\documentclass[11pt]{article}

\usepackage{natbib}

\usepackage{graphicx}
\usepackage{here}
\usepackage[labelfont=bf,hang]{caption}
\usepackage[labelfont=sf]{subfig}
\usepackage{color}
\usepackage{bm}
\usepackage{amsmath, amsthm, amsfonts}

\usepackage{cases}
\usepackage{newtxtext}
\usepackage{fancyhdr}
\usepackage{longtable}
\usepackage{here}

\usepackage{lineno}
\makeatletter
	\let\start@align@nopar\start@align
	\let\start@gather@nopar\start@gather
	\let\start@multline@nopar\start@multline
	\let\start@numcases@nopar\start@numcases
	\long\def\start@align{\par\vspace{-1em}\start@align@nopar}
	\long\def\start@gather{\par\vspace{-1em}\start@gather@nopar}
	\long\def\start@multline{\par\vspace{-1em}\start@multline@nopar}
	\long\def\start@numcases{\par\vspace{-1em}\start@numcases@nopar}
\makeatother

\makeatletter
	\def\appendixname{Appendix~}
	\renewcommand\appendix{\par
	  \setcounter{section}{0}%
	  \setcounter{subsection}{0}%
	  \setcounter{equation}{0}
	  \gdef\thefigure{\@Alph\c@section.\arabic{figure}}%
	  \gdef\thetable{\@Alph\c@section.\arabic{table}}%
	  \gdef\thesection{\appendixname\@Alph\c@section}%
	  \gdef\thesubsection{\@Alph\c@section.\arabic{subsection}}%
	  \@addtoreset{equation}{section}%
	  \gdef\theequation{\@Alph\c@section.\arabic{equation}}%
	  \addtocontents{toc}{\string\let\string\numberline\string\tmptocnumberline}{}{}
	}
\makeatother

\usepackage{enumitem}
	\setlist[itemize]{noitemsep}
	\setlist[description]{noitemsep}
	\setlist[enumerate]{noitemsep, font=\bf}

\usepackage{endnotes}

\usepackage[top=30truemm,bottom=30truemm,left=30truemm,right=30truemm]{geometry}

\usepackage[
	bookmarks=false,
	colorlinks=true,
	citecolor=blue,
	urlcolor=blue,
	linkcolor=blue,
 	pdfstartview=FitH,
]{hyperref}

\usepackage[capitalise]{cleveref}
\crefname{theo}{Theorem}{Theorems}
\crefname{prop}{Proposition}{Proposition}
\crefname{eq}{Eq.}{Eqs.}
\crefname{fig}{Fig.}{Figs.}
\crefname{tab}{Tab.}{Tabs.}
\crefname{sec}{Section}{Sections}

\def\({\left(}
\def\){\right)}
\def\[{\left[}
\def\]{\right]}
\def\<{\left\langle}
\def\>{\right\rangle}
\def\<<{\left\{}
\def\>>{\right\}}
\def\ds{\displaystyle}
\def\dif{{\rm d}}

\def\->{$\rightarrow$}

\allowdisplaybreaks

\reversemarginpar

\begin{document}


\title{Fundamental diagram of urban rail transit considering train--passenger interaction\footnote{This paper is an enhanced version of \citet{Seo2017train_arxiv, seo2017trainfd_trb}. Submitted to an international journal.}}
\author{
	Toru Seo\footnote{Tokyo Institute of Technology, 2-12-1-M1-13 O-okayama, Meguro, Tokyo 152-8552, Japan; The corresponding author; seo.t.aa@m.titech.ac.jp}	\quad
	Kentaro Wada\footnote{University of Tsukuba, 3F1211, 1-1-1 Tennodai, Tsukuba, Ibaraki 305-8573, Japan}	\quad
	Daisuke Fukuda\footnote{The University of Tokyo, 7-3-1 Hongo, Bunkyo-ku, Tokyo 113-8656, Japan}
}

\date{\today}
\maketitle

\begin{abstract}
	Urban rail transit often operates with high service frequencies to serve heavy passenger demand during rush hours.
	Such operations can be delayed by two types of congestion: train congestion and passenger congestion, both of which interact with each other.
	This delay is problematic for many transit systems, since it can be amplified due to the interaction.
	However, there are no tractable models describing them; and it makes difficult to analyze management strategies of congested transit systems in general and tractable ways.
	To fill this gap, this article proposes simple yet physical and dynamic model of urban rail transit.
	First, a fundamental diagram of transit system (i.e., theoretical relation among train-flow, train-density, and passenger-flow) is analytically derived considering the aforementioned physical interaction.
	Then, a macroscopic model of transit system for dynamic transit assignment is developed based on the fundamental diagram.
	Finally, accuracy of the macroscopic model is investigated by comparing to microscopic simulation.
	The proposed models would be useful for mathematical analysis on management strategies of urban rail transit systems, in a similar way that the macroscopic fundamental diagram of urban traffic did.
\end{abstract}

\begin{center}
\end{center}

\section{Introduction}\label{sec_intro}

Urban rail transit systems such as metro is handling significant transportation needs of metropolitan areas \citep{vuchic2005train}.
Its most notable usage is the morning commute, in which heavy passenger demand is concentrated in a short time period.
It is known that such transit systems often suffer from delays caused by congestion, even if no serious incidents or accidents occur \citep{kato2012train,tirachini2013train,kariyazaki2015train}.
Therefore, appropriate management of transit systems is required; especially, travel demand management for mass transit systems has been gaining attention recently \citep{Halvorsen2019train, Huan2021tdm}.

One of the approaches to find management strategies of transit systems is theoretical analysis with simplifications, such as use of certain static models with constant travel time \citep{decea1993train, tabuchi1993train, Kraus2002train, tian2007train, Gonzales2012train, Trozzi2013train, depalma2015train, depalma2015trainb}.
In this approach, general policy implications can be obtained thanks to the simplicity and tractability of the analysis.
However, they may not be sufficient to investigate dynamic operation and demand management strategies.

In congested transit systems, dynamical interaction among trains and passengers plays essential roles to determine the system's operational behavior, and the travel time can be dynamically and significantly changed due to this interaction.
For instances, there are two types of congestion in transit systems:
\begin{itemize}[nosep]
 	\item {\it train-congestion}: congestion involving consecutive trains using the same tracks,
 	\item {\it passenger-congestion}: congestion of passengers who are boarding to a train, namely, bottleneck congestion at the doors of a train while it is stopped at a station \citep{lam1998train, wada2012train_en, kariyazaki2015train},\footnotemark{}
\end{itemize}
\footnotetext{
	Note that passenger-congestion differs from in-vehicle {\it passenger-crowding} \citep{Kumagai2020train}, which results in discomfort due to standing and crowding, but is not necessarily cause any delay directly.
}
and these two types of congestion interact with each other and cause delay \citep{newell1964maintaining, Kusakabe2010train, wada2012train_en,kato2012train,tirachini2013train,kariyazaki2015train, Cuniasse2015train}.
The most typical phenomena involving the dynamic train--passenger interaction would be the ``knock-on delay'' \citep{carey1994train}---this is a train equivalent of the ``bus bunching'' \citep{newell1964maintaining,Daganzo2009bus}.
For example, assume that passenger-congestion happened temporally due to high demand.
It would extend the dwelling time of a train at a station.
Then, this extended dwell time would interrupt the operation of subsequent trains, and cause train-congestion on the track.
It would deteriorate the passenger throughput, and thus the passenger-congestion at stations would intensify.
This kind of dynamical phenomena cannot be captured by static models.

\begin{figure}[tbp]
	\centering
	\subfloat[Tokyu Den-en-toshi Line, Tokyo, Japan]{\includegraphics[width=.49\hsize]{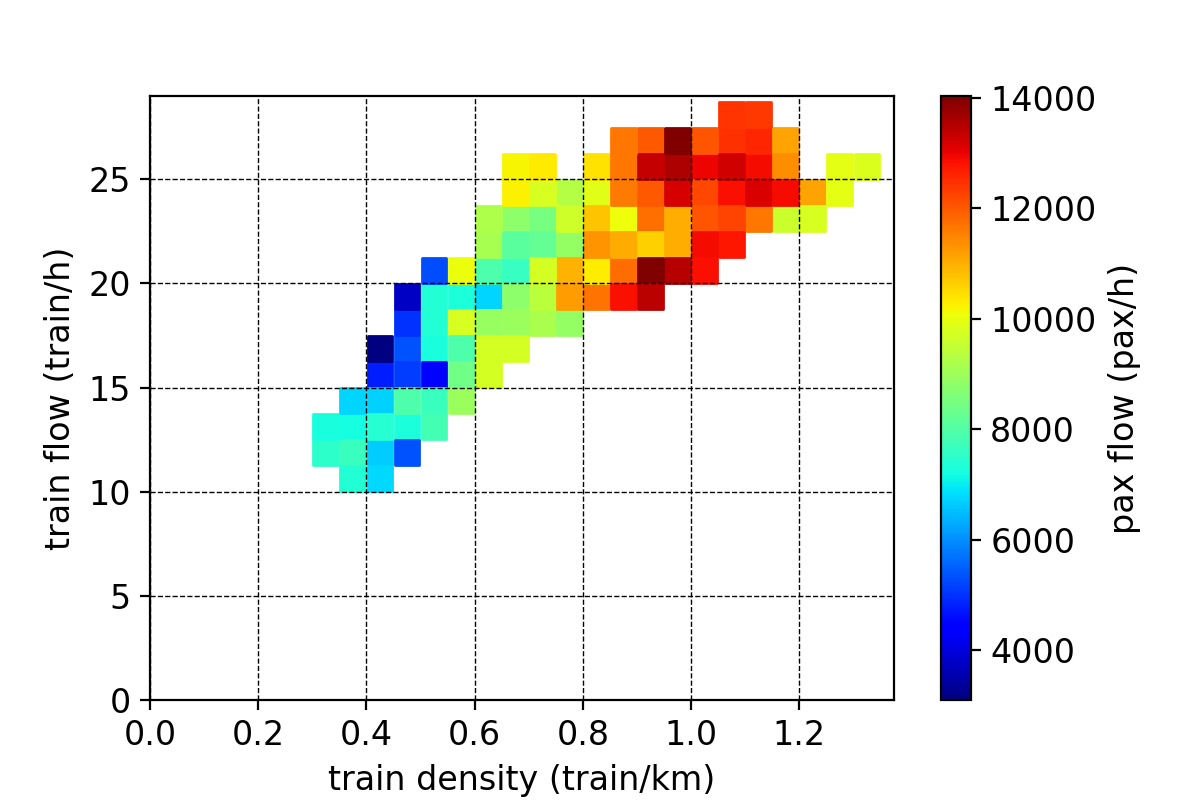}\label{fd_Tokyo}}
	\subfloat[Boston Subway Red Line, Massachusetts, US]{\includegraphics[width=.49\hsize]{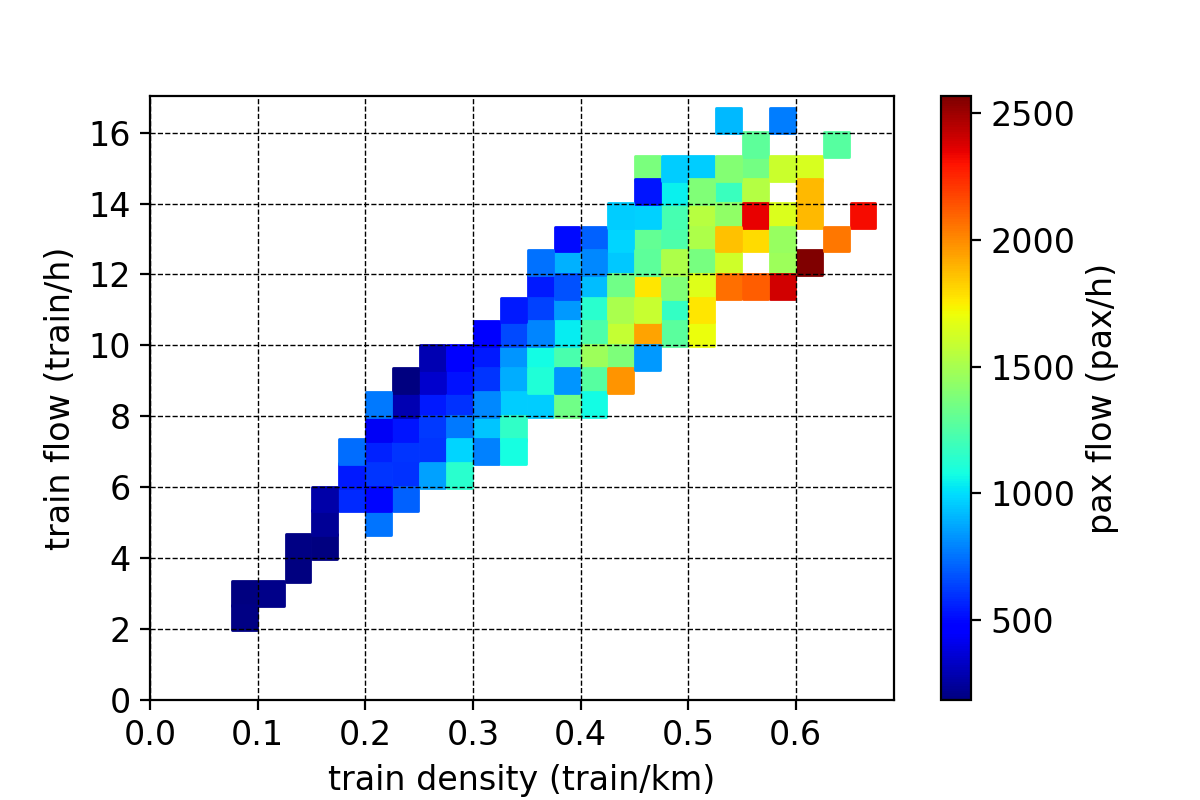}\label{fd_Boston}}
	\caption{Observed 3-dimensional relations among train-flow, train-density, and passenger-flow. The $x$-coordinates represent the train-flow (i.e., the number of trains per a kilometer), the $y$-coordinates represent the train-flow (i.e., the number of passing trains per an hour), the slope from the origin to an arbitrary point indicates the train-speed (i.e., the average speed of trains) of that point, and the color of each point represents the passenger-flow (i.e., average demand of passengers per an hour) of that point. ``pax'' in the figure is an abbreviation for ``passenger''. Data sources: \citet{Fukuda2019trainfd}, \citet{zhang2019trainfd}, Tokyu Railways, Massachusetts Bay Transportation Authority.}
	\label{fd_real}
\end{figure}

A consequence of such dynamical passenger--train interaction can be found in macroscopic states of transit systems.
\cref{fd_real} shows observed 3-dimensional relations among states of transit systems, that is, train-flow (train/h), train-density (train/km), and passenger-flow (passenger/h).
The visualization is based on the concepts of the fundamental diagram \citep{Greenshields1935fd} of urban traffic.
Although \cref{fd_Tokyo} and \cref{fd_Boston} show data from completely different transit systems, they have remarkable similarities.
First, as the passenger-flow increases, the train-density increases; this could be a result of transit operators responding to increased passenger-demand.
Second, as the passenger-flow increases, the average service speed of trains and train-flow decrease; this could be a result of the aforementioned congestion due to the interaction among trains and passengers.
It would be preferable if we have a theoretical model of this phenomena, because it would be useful to obtain general principles on transit operations; however, to our knowledge, such a model does not exist in the literature.\footnotemark{}
\footnotetext{
	Several detailed operation models have been proposed to capture the detailed mechanism of the dynamics of interaction \citep[see][and references therein]{vuchic2005train, Koutsopoulos2007train, Parbo2016train, Cats2016transit, Li2017train, Alonso2017train, Cunha2021train}, and these have been used to develop efficient operation schemes. 
	However, these models are often based on microscopic simulation, and thus their purposes tend to be case-specific optimization and evaluation.
	It would be difficult to use them to derive the relation depicted in Fig.~\ref{fd_real} or to obtain general policy implications for management strategies, as they are essentially complex and intractable.
}

This study derives a theoretical relation among the state variables of transit systems similar to \cref{fd_real} based on the microscopic operation principles.
It is modeled as a fundamental diagram (FD), which is a well-known concept in vehicular traffic flow theory.
The original FD describes relation between vehicular flow and density, and it can be used to describe dynamic evolution of traffic by combining with other principles in a tractable manner \citep{Lighthill1955LWR, Richards1956LWR, mahmassani1984nfd, Geroliminis2007mfd}.
In fact, several recent studies have employed FDs of transit systems to describe train-congestion by modeling the relation between train-flow and train-density \citep{Cuniasse2015train, Corman2019train, de2021train}.
The novel feature of this study is the incorporation of passengers in an analytical way.
Furthermore, this study develops a dynamic transit assignment method based on the proposed FD.

This study proposes tractable models of the dynamics of urban rail transit considering the physical interaction between train-congestion and passenger-congestion.
In Section \ref{sec_ass}, a microscopic model of a rail transit system is introduced based on a passenger boarding model and a train cruising model.
In Section \ref{sec_fd}, the operation performance of the microscopic model is analyzed.
Specifically, a mathematically tractable relation among train-flow, train-density, and passenger-flow is derived---that is, a {\it fundamental diagram (FD)}.
The model can be also viewed as a variation of 3-dimensional macroscopic fundamental diagram \citep{mahmassani1984nfd, Geroliminis2007mfd} with an analytical derivation.
This is the key contribution of this study.
In Section \ref{sec_dynamic}, a macroscopic loading model of a transit system is developed based on the proposed FD.
The model describes the aggregated behavior of trains and passengers in a urban-scale spatial domain based on the FD.
In Section \ref{sec_valid}, the approximation accuracy and other properties of the proposed macroscopic model are investigated through a comparison with microscopic simulation.
Section \ref{sec_conc} concludes this article.
Note that empirical validation of the proposed model based on actual data is out of scope of this study.
Such validation is now being conducted by some of the authors and preliminary results that support the model have been obtained \citep{Fukuda2019trainfd, zhang2019trainfd}.

\section{Microscopic Model of Rail Transit System}\label{sec_ass}

This section introduces a microscopic model of rail transit system, from which we derive the FD in Section \ref{sec_fd}.
It consists of two microscopic operation principles, namely, a {\it passenger boarding model} which describes the train's dwell behavior at a station for passenger boarding and a {\it train cruising model} which describes the cruising behavior on the railroad.
This microscopic model has been proposed by \citet{wada2012train_en} to analyze train bunching. 

\subsection{Rail Transit Operation Principles}\label{sec_ass2}

\begin{figure}[tbp]
	\centering
	\includegraphics[width=0.7\hsize]{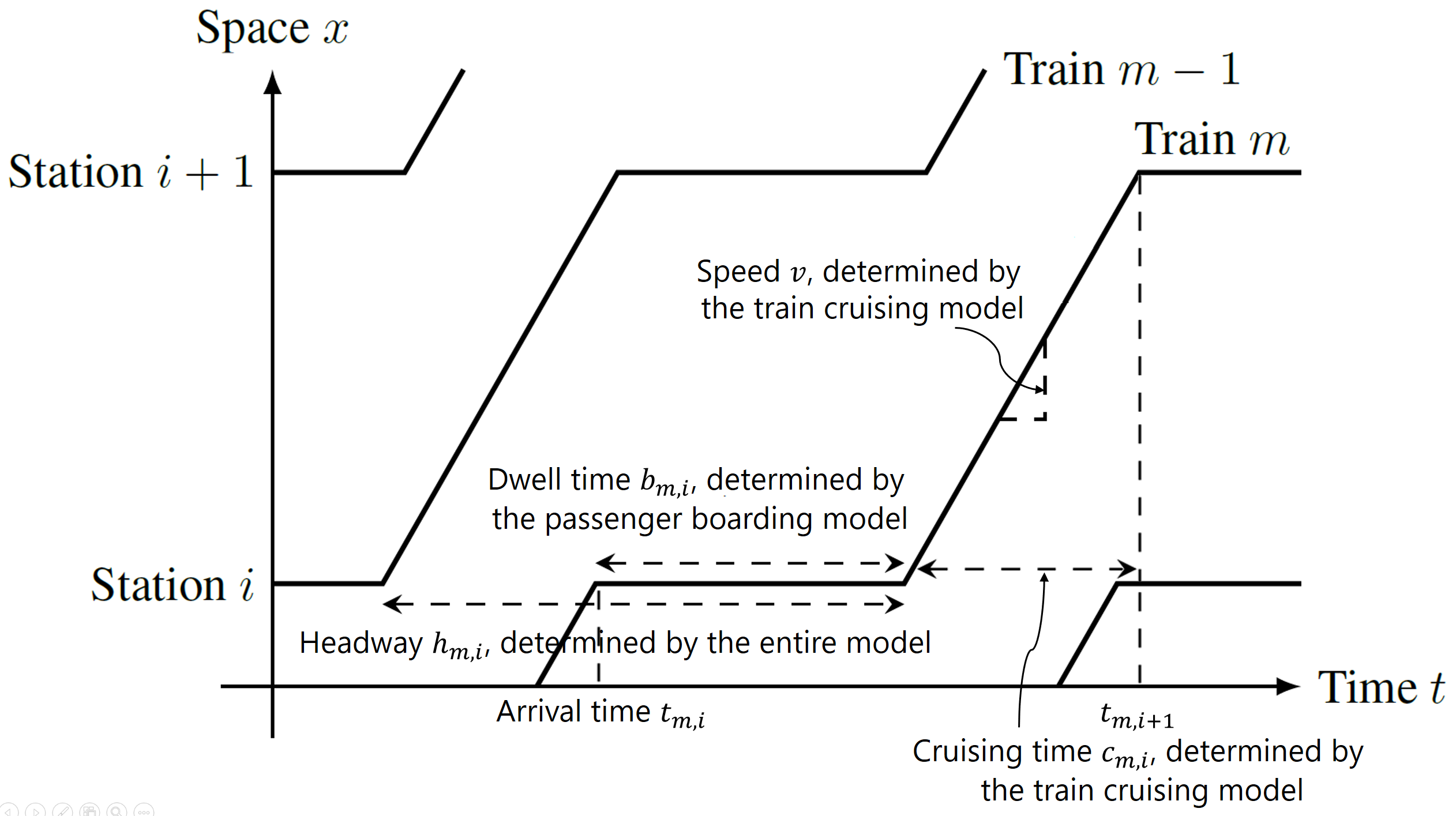}
	\caption{Illustration of the microscopic model of rail transit system.}
	\label{tsd_anoted}
\end{figure}

Consider a railway system on a single line track, where trains and stations are indexed by $m$ and $i$, respectively. 
We assume that all trains stop at every station.
Let $t_{m,i}$ be the arrival time of train $m$ at station $i$. 
Then, a dynamical system that represents each train motion is given by
\begin{align}
	& t_{m,i+1} = t_{m,i} + b_{m,i} + c_{m,i} & \forall m, i \label{eq:micro}
\end{align}
where $b_{m,i}$ is the passenger boarding time of train $m$ at station $i$, and $c_{m,i}$ is the trip time of train $m$ between stations $i$ and $i + 1$, which are determined by the two operational submodels (see Figure \ref{tsd_anoted}). 

The passenger boarding time is modeled using a queuing model.
That is, the flow-rate of passenger boarding is assumed to be constant, $\mu_p$; and there is a buffer time (e.g., time required for door opening/closing), $g_b$, for the dwell time. 
Then, the dwell time of a train at a station, $b_{m,i}$, is represented as
\begin{align}
	b_{m,i} = \frac{q_{p,i}h_{m,i}}{\mu_p} + g_{b,i},	\label{boarding}
\end{align}
where $q_{p,i}$ is the (possibly time-dependent) passenger demand flow rate at station $i$, $h_{m,i}\equiv t_{m,i} - t_{m-1,i}$ is the time-headway, and thus $q_{p,i}h_{m,i}$ is the number of waiting passengers at the station.\footnotemark{}
This can be considered as a special case of \citet{lam1998train}.
All passengers waiting a train at a station are assumed to board the first train arrived.

\footnotetext{
	\label{note_paxflow}
	In reality, there are passengers alighting a train, in addition to ones boarding.
	By carefully distinguishing the two types of passengers and replacing the terminology in the main text, the discussions in the main text are valid and the final results are not altered.
	For example, ``the number of boarding passengers'' can be replaced with ``the sum of the number of boarding passengers and the number of alighting passengers''.
	However, it will complicate the discussions; therefore, we ignore passengers alighting a train.
}%

The cruising behavior of a train is modeled using the Newell's simplified car-following model \citep{newell2002carfollowing}.\footnotemark{}
\footnotetext{%
	Newell's simplified car-following model is a special case of the well-known road traffic flow model, the Lighthill--Whitham--Richards (LWR) model \citep{Lighthill1955LWR, Richards1956LWR, Newell1993kinematic}.
	Although the LWR model is known as a ``macroscopic'' model based on continuum fluid approximation, \citet{newell2002carfollowing} showed that it is equivalent to a microscopic car-following model proposed by his paper.
}%
In this model, a train travels by  maintaining the minimum safety clearance.
Specifically, $x_{m,i}(t)$, position of a train $m$ between stations $i$ and $i+1$ at time $t$, is described as
\begin{align}
	x_{m,i}(t) = \min\left\{x_{m,i}(t-\tau) + v \tau, x_{m-1,i}(t-\tau) - \delta\right\},		\label{newellx}
\end{align}
where $m-1$ indicates the preceding train of train $m$, $\tau$ is the physical minimum time-headway, $v$ is the desired cruising speed that is determined by a (fixed-block or moving-block) signal control system, and $\delta$ is the minimum spacing.

We call the traffic is in {\it free-flowing regime} if the train travels between stations at the free-flow speed $v_{f}$ (the maximum speed of the track), i.e., the train motion is represented by the first term with $v = v_{f}$ in the minimum operation of Eq.~\eqref{newellx}. 
We call the traffic is in {\it congested regime} if the train is required to decrease its speed to maintain both the safety headway $\tau$ and distance $\delta$.
In this case, the second term of Eq.~\eqref{newellx} is active.
The speed profile in this regime may differ in different train operators (i.e., signal control systems) and drivers. 
We employ one of the simplest approximations of this speed profile, that is, the train travels between stations at a constant speed while maintaining the minimum safety clearance. 

%
%

\subsection{Validity of Assumptions}

In this model, the dwell time of a train is determined by the number of boarding passengers, not by a pre-determined timetable.
Although this seems like inconsistency between the proposed model and actual schedule-based train operations, this can be considered as a reasonable approximation of average operation pattern of schedule-based train operations.
The reasons are as follows.
First, in a congested urban areas, it is common that passenger boarding time is not negligible and occasionally delay transit operation, as reviewed in Section \ref{sec_intro}.
Therefore, in order to maintain a scheduled operation based on a timetable, this timetable has to be determined considering the passenger demand \citep[e.g.,][]{niu2013timetable}.
Consequently, the dwelling time in such timetable can be considered as similar to the proposed passenger boarding model \eqref{boarding} where $q_{p,i}h_{m,i}$ is interpreted as an average number of waiting passenger and $g_b$ is interpreted as a buffer time to deal with fluctuation of the demand.

Second, the passenger boarding model with a constant capacity is consistent with the modelling of ordinary pedestrian flows for a fixed-width bottleneck \citep{lam1998train, hoogendoorn2005pedestrian}.
Meanwhile, there is no stock capacity for passengers in the presented model; in other words, a train can transport infinite number of passengers.
This is a limitation of the current model; however, unless the passenger demand is excessive level (e.g., where not all of the waiting passenger can board an arriving train), this limitation will not be problematic.

The train cruising model \eqref{newellx} can be considered as a lower order but a reasonable approximation of the train movement in the sense that the fundamental operating principle is consistent with practical controls and existing studies \citep{carey1994train, higgins1998train, Huisman2005train}. 
That is, each train has to maintain a headway and a spacing that are greater than the given minimum ones. 
The model directly corresponds to the ``moving block control,'' \citep{Dicembre2011train} but it can be also viewed as an approximation of traditional ``fixed block control\footnotemark{}.''
\footnotetext{
	The equivalent cellular automaton models of Newell's car-following model \citep{daganzo2006kinematic} may represent the fixed block control as in the existing studies \citep[e.g.,][]{Li2005train,kariyazaki2015train}.
}  
The main assumption here is the constant speed assumption in the congested regime. 
This approximates a train operation with low acceleration rates for ensuring the comfort of passengers and less energy loss. 
In addition, this assumption can describe the congested situation where the train stops between stations, on average, as we will show in the numerical experiment in Section \ref{sec_valid}.

\section{Fundamental Diagram of Rail Transit System}\label{sec_fd}

In this section, we derive an FD of a rail transit system described by the microscopic model formulated in Section \ref{sec_ass}.
The FD is defined as the relation among train-flow, train-density, and passenger-flow.

\subsection{Steady State of Rail Transit System}\label{sec_steady}

We consider the {\it steady state} of the proposed microscopic model.
The steady state is an idealized traffic state that does not change over time, and its traffic state variables (typically combination of flow, density, and speed) are characterized by special relation (called an FD) of the traffic flow \citep{daganzo1997book}.
Let us consider a homogeneous rail transit system in which the stations and passenger demands are homogeneously distributed over the line, i.e., $l_{i} = l$, $q_{p,i} = q_{p}$ and other parameters ($g_{b}, v_{f}$, $\tau$, and $\delta$) are the same for all stations and trains. 
Then, the steady state is defined as a state that, for a given steady passenger demand $q_{p}$, the time-headway between successive trains, $H$, is time-independent. 
Note that $q_p < \mu_p$ must be satisfied; otherwise, passenger boarding will never end.

\begin{figure}[hbtp]
	\centering
	
	\subfloat[Free-flowing regime: $v = v_f$, $h_f > 0$.]{
		\includegraphics[height=14em]{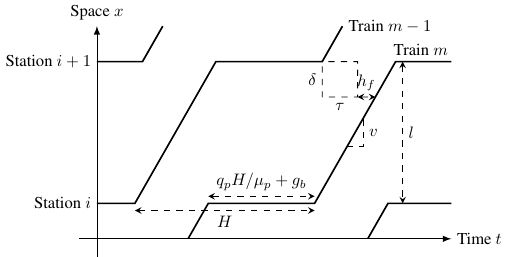}
	\label{tsd_free}}\\
	\subfloat[Critical regime: $v = v_f$, $h_f = 0$.]{
		\includegraphics[height=14em]{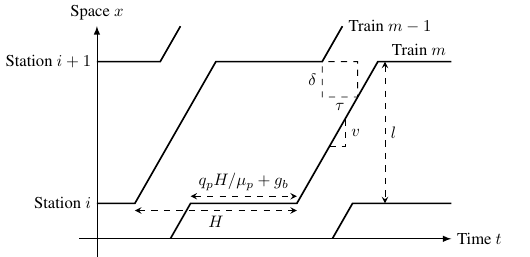}
	\label{tsd_crit}}\\
	\subfloat[Congested regime: $v < v_f$, $h_f = 0$.]{
		\includegraphics[height=14em]{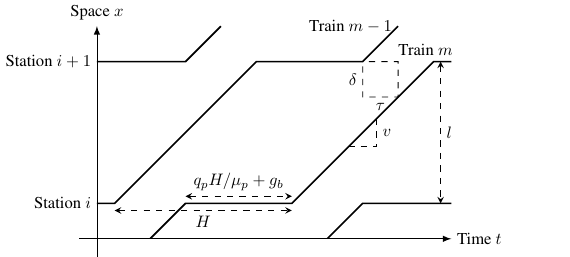}
	\label{tsd_cong}}
	
	\caption{Time--space diagrams of rail transit system under steady states.}
	\label{tsd}
\end{figure}

Transit systems under different steady states are illustrated as time--space diagrams in Fig.~\ref{tsd}.
In each sub-figure, train $m$ arrives at and departs from station $i$, then travels to station $i+1$ at cruising speed $v$, and finally arrives at station $i+1$.
The main differences between each sub-figure are density of trains and, consequently, traffic regime.
In Fig.~\ref{tsd_free}, the density is small so that the speed $v$ is equal to the free-flow speed $v_f$ and $h_f$ is greater than zero; therefore the state is classified into the free-flowing regime.
In Fig.~\ref{tsd_crit}, the density is medium so that the speed is equal to $v_f$ and $h_f$ is equal to zero; therefore, the state is classified into the critical regime.
In Fig.~\ref{tsd_cong}, the density is large so that the speed is less than $v_f$; therefore, the state is classified into the congested regime.

\subsection{Fundamental Diagram}

In general, the followings are considered as the traffic state variables of a rail transit system:
\begin{itemize}[noitemsep]
	\item train-flow $q$,
	\item train-density $k$, 
	\item train-mean-speed $\bar{v}$, 
	\item passenger-flow $q_p$, 
	\item passenger-density $k_p$, 
	\item passenger-mean-speed $\bar{v}_p$.
\end{itemize}
Among these, there are three independent variables: for example, the combination of $q$, $k$, and $q_p$.
This is because of the identities $q=k\bar{v}$ and $q_p=k_p \bar{v}_p$, and $\bar{v}=\bar{v}_p$.\footnotemark{}
\footnotetext{
	Note that the mean speed $\bar{v}$ differs from the cruising speed $v$; the former takes the dwelling time at a station and cruising between stations into account, whereas the latter only considers the cruising time.
}%

Now suppose that the relation among the independent variables of the traffic state under every steady state is expressed using a function $Q$ as
\begin{align}
	q=Q(k,q_p).	\label{fdeq0}
\end{align}
The function $Q$ is regarded as an FD of the rail transit system.
In fact, by assuming that the rail transit operation principle follows Eqs.~\eqref{boarding} and \eqref{newellx}, the FD function is analytically derived as
\begin{align}
	Q(k,q_p)= \left\{
		\begin{array}{ll}
			\ds\frac{lk - q_p/\mu_p}{g_b+l/v_f},	&	\mathrm{if}~	k < k^*(q_p),\\[1em]
			- \ds\frac{l\delta}{(l-\delta)g_b+\tau l} (k-k^*(q_p)) + q^*(q_p), &	\mathrm{if}~ k \geq k^*(q_p),
		\end{array}
	\right.	\label{fdeq}
\end{align}
with
\begin{align}
	&q^*(q_p) = \frac{1-q_p/\mu_p}{g_b+\delta/v_f+\tau},	\label{critq}\\
	&k^*(q_p) = -\frac{(l-\delta)/v_f-\tau}{(g_b+\delta/v_f+\tau) \mu_p l} q_p + \frac{g_b+l/v_f}{(g_b+\delta/v_f+\tau) l},	\label{critk}
\end{align}
where $q^*(q_p)$ and $k^*(q_p)$ represent train-flow and train-density, respectively, at a critical state with passenger-flow $q_p$.
For the derivation, see \ref{apndx_fd}.
Although the FD equations \eqref{fdeq}--\eqref{critk} look complicated, they represent a simple relation: a piecewise linear (i.e., triangular) relation between $q$ and $k$ under fixed $q_p$.
See Fig.~\ref{fdexample} for a numerical example of the FD which we will explain later.

\subsection{Discussions}

The FD is interpreted as a function that describes transit operation performance (train-flow $q$, headway $H=1/q$, and mean-speed $\bar{v}=q/k$) under a given train supply (train-density $k$) and passenger demand (passenger-flow $q_p$) for the given technical parameters of the transit system ($\mu_p, g_b, v_f, \tau, \delta, l$).
Therefore, it can be considered as a similar concept to the macroscopic fundamental diagram (MFD) \citep{Geroliminis2007mfd, Daganzo2007gridlock}, which describes a road network throughput under a given number of vehicles and technical parameters of the road network. 

\subsubsection{Numerical Example}\label{sec_FD_num}

\begin{figure}[htbp]
	\centering
	\includegraphics[width=0.7\hsize]{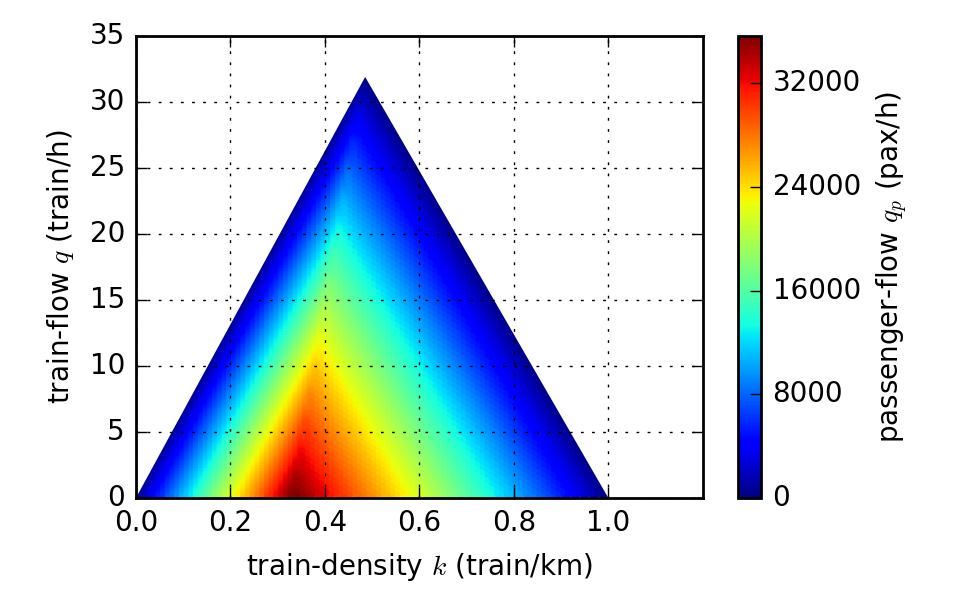}
	\caption{Numerical example of the FD.}
	\label{fdexample}
\end{figure}

First of all, for ease of understanding, we show a numerical example of the FD in Fig.~\ref{fdexample}.
The parameter values are presented in Table \ref{fdparameters}.
In the figure, the horizontal axis represents train-density $k$, the vertical axis represents train-flow $q$, and the plot color represents passenger-flow $q_p$.
The slope of the straight line from a traffic state to the origin represents the mean speed $\bar{v}$ of the state.

\begin{table}[hbtp]
	\centering
	\caption{Parameters of the numerical example.}
	\label{fdparameters}
	\footnotesize
	\begin{tabular}{rl}
		\hline
		parameter & value	\\
		\hline
		$u$ & 70 km/h	\\
		$\tau$ & 1/70 h	\\
		$\delta$ & 1 km	\\
		$\mu_p$ & 	36000 pax/h	\\
		$g_b$ & 10/3600 h	\\
		$l$ & 3 km	\\
		\hline
	\end{tabular}
\end{table}

For example, the figure can be read as follows.
Suppose that the passenger demand per station is $q_p=16000$ (pax/h).
If the number of trains in the transit system is given by the train-density $k=0.3$ (train/km), then the resulting train traffic has a train-flow of $q \simeq 15$ (train/h) and a mean speed of $\bar{v} \simeq 50$ (km/h).
This is the traffic state in the free-flowing regime.
There is a congested state corresponding to a free-flowing state: for the aforementioned state with ($q$, $k$, $\bar{v}$) $\simeq$ (15 veh/h, 0.3 veh/km, 50 km/h), the corresponding congested state is (15 veh/h, 0.55 veh/km, 27 km/h).
The critical state under $q_p=16000$ (pax/h) is (22 veh/h, 0.42 veh/km, 52 km/h).
Notice that this state has the fastest mean speed under the given passenger demand.
The triangular $q$--$k$ relation mentioned before is clearly shown in the figure; the ``left edge'' of the triangle corresponds to the free-flowing regime, the ``top vertex'' corresponds to the critical regime, and the ``right edge'' corresponds to the congested regime.

By comparing the theoretical FD (\cref{fdexample}) with the actual data (\cref{fd_real}), some similarities can be found.
The two features found in the actual data (as the passenger-flow increases, the train-density increases; and as the passenger-flow increases, the train-speed and train-flow decrease) can be interpreted that the actual data are from a part of free-flow regime of the theoretical FD.
Furthermore, in the high train-density regime in the Tokyo data (\cref{fd_Tokyo}), we observed a slight drop of passenger-flow; this might be a congested regime of the theoretical FD.
From these results, we can say that the theoretical model explains the actual data to some extent.

\subsubsection{Detailed Features of Fundamental Diagram}\label{sec_FD_feature}

The FD has the following theoretical features which are analytically derived from Eq.~\eqref{fdeq}.
They can easily be found in the numerical example in Fig.~\ref{fdexample}.

As mentioned, the traffic state of a transit system is categorized into three regimes (free-flowing, critical, and congested), as in the standard traffic flow theory.
Therefore, there is a critical train-density $k^*(q_p)$ for any given $q_p$.
Train traffic is in the free-flowing regime if $k < k^*(q_p)$, in the critical regime if $k = k^*(q_p)$, or in the congested regime otherwise.
The congested regime can be considered as inefficient compared with the free-flowing regime, because the congested regime takes more time to transport the same volume of passengers.
The critical regime is the most efficient in the sense that its travel time (i.e., $1/\bar{v}$, $1/\bar{v}_p$) and in-vehicle crowding (i.e., number of passengers per train, $q_p/q$) are minimum under a given passenger demand.
However, the critical regime requires more trains (i.e., higher train-density) than the free-flowing regime; therefore, it may not be the most efficient if the operation cost is taken into account.

Even in the critical regime, the mean speed $\bar{v}$ is inversely proportional to passenger demand $q_p$.
This means that travel time increases as passenger demand increases.
In addition, the size of the feasible area of $(q,k)$ narrows as $q_p$ increases.
Thus, the operational flexibility of the transit system declines as the passenger demand increases.

Flow and density of trains in the critical regime satisfy the following relations:
\begin{align}
	q^*(q_p) =& \frac{l}{(l-\delta)/v_f-\tau} k^*(q_p) - \frac{1}{(l-\delta)/v_f-\tau}.	\label{criteq}
\end{align}
(Here, we have assumed $(l-\delta)/v_f-\tau \neq 0$.)
Therefore, the critical regime is represented as a straight line whose slope ($l/[(l-\delta)/v_f-\tau]$) is either positive or negative in the $q$--$k$ plane.
This implies a qualitative difference between transit systems.
Specifically, if the slope is positive, a transit operation with constant train-density would transition from the free-flowing regime to the congested regime as passenger demand increases (Fig.~\ref{fdexample}).
On the contrary, if the slope is negative, such an operation would transition from free-flowing to congested as passenger demand {\it decreases}.
This seems paradoxical, but it is actually reasonable because the operational efficiency can be degraded if the number of trains is excessive compared to passenger demand.

The FD describes an transit system's performance under a steady state operation as mentioned.
Under the presence of well-designed adaptive control strategies, such as schedule-based and headway-based control \citep{Daganzo2009bus, wada2012train_en}, the steady state is likely to be realized.
This is because the aim of such adaptive control is usually to eliminate bunching---in other words, such control makes the operation steady.
Therefore, it can be expected that the FD could be useful to describe average performance of actual transit system, which is usually not steady due to heterogeneity among passenger demand and train supply.
This issue is numerically validated in Section \ref{sec_valid}.

Last but not least, it is worth mentioning that all parameters in the proposed model have an explicit physical meaning.
Therefore, the parameter calibration required to approximate an actual transit system is relatively easy.

\subsubsection{Relation to the Macroscopic Fundamental Diagram}\label{sec_mfd}

The proposed FD resembles the MFD \citep{Geroliminis2007mfd, Daganzo2007gridlock} and its extensions \citep[e.g.,][]{Geroliminis2014mfd, Chiabaut2015mfd} as mentioned.
They are similar in the following sense.
First, they both consider dynamic traffic.
Second, they both describe the relations among macroscopic traffic state variables in which the traffic is not necessarily steady or homogeneous at the local scale (i.e., they use area-wide aggregations based on Edie's definition; see \ref{apndx_edie}).
Third, they both have unimodal relations, meaning that there are free-flowing and congested regimes, where the former has higher performance than the latter; in addition, there is a critical regime where the throughput is maximized.
Therefore, it is expected that existing approaches for MFD applications, such as modeling, control and the optimization of transport systems \citep[e.g.,][]{Daganzo2007gridlock, geroliminis2009mfd, geroliminis2013mfd, Fosgerau2015bathtub}, are also suitable for the proposed transit FD.

However, there are substantial differences between the proposed FD and the existing MFD-like concepts.
In comparison with the original MFD \citep{Geroliminis2007mfd, Daganzo2007gridlock} and its railway variant \citep{Cuniasse2015train}, the proposed FD has an additional dimension, that is, passenger-flow.
In comparison with the three-dimensional MFD of \cite{Geroliminis2014mfd}, which describes the relations among total traffic flow, car density, and bus density in a multi-modal traffic network, the proposed FD explicitly models the physical interaction among the three variables.
In comparison with the passenger MFD of \cite{Chiabaut2015mfd}, which describes the relation between passenger flow and passenger density when passengers can choose to travel by car or bus, in the proposed FD, passenger demand can degrade the performance (i.e., speed) of the vehicles because of the inclusion of the boarding time.

\section{Dynamic Model Based on Fundamental Diagram}\label{sec_dynamic}

Recall that the proposed FD describes the relationship among traffic variables under the steady state.
It means that the behavior of a dynamical system in which demand and supply change over time is not described solely by the FD.
This feature is the same as in the road traffic FD and MFDs.
In this section, we formulate a model of urban rail transit operation where the demand (i.e., passenger-flow) and supply (i.e., train-density) change dynamically.
In this proposed model, individual train and passenger trajectories are not explicitly described; therefore, the model is called macroscopic.

\begin{figure}[hbtp]
	\centering
	\includegraphics[width=0.99\hsize]{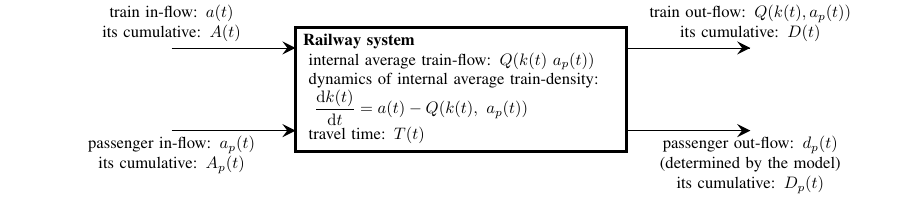}
	\caption{Railway system as an input--output system.}
	\label{inputoutput}
\end{figure}

The proposed model is based on an {\it exit-flow model} \citep{Merchant1978dta, carey2004exitflow} in which the proposed FD is employed as the exit-flow function.
Specifically, the transit system is considered as an input--output system, as illustrated in Fig.~\ref{inputoutput}.
The exit-flow modeling approach is often employed for area-wide traffic approximations and analysis using MFDs, such as optimal control to avoid congestion \citep{Daganzo2007gridlock} and analyses of user equilibrium and social optimum in morning commute problems \citep{geroliminis2009mfd}.
The advantage of this approach is that it would be possible to conduct mathematically tractable analysis of dynamic, large-scale, and complex transportation systems, where the detailed traffic dynamics are difficult to model in a tractable manner---this is the case for transit operations.

\subsection{Formulation}\label{sec_dynamic_formulation}

Let $a(t)$ be the inflow of trains to the transit system, $a_p(t)$ be the inflow of passengers, $d(t)$ be the outflow of trains from the transit system, and $d_p(t)$ be the outflow of passengers, on time $t$ respectively.
We set the initial time to be $0$.
Let $A(t)$, $A_p(t)$, $D(t)$, and $D_p(t)$ be the cumulative values of $a(t)$, $a_p(t)$, $d(t)$, and $d_p(t)$, respectively (e.g., $A(t)=\int_0^t a(s) \dif s$).
Let $T(t)$ be the travel time of a train that entered the system at time $t$, and let its initial value $T(0)$ be given by the free-flow travel time under $q=a(0)$ and $q_p=a_p(0)$.
To simplify the formulation, the trip length of the passengers is assumed to be equal to that of the trains.\footnotemark{}
\footnotetext{
	This assumption is reasonable if the average trip length is shared by trains and passengers.
	If they are different, a modification such as $T_p(t) = T(t)/\lambda$, where $\lambda$ is the ratio of average trip length of the passengers to that of the trains, would be possible.
}%
This means that $T(\cdot)$ is the travel time of both the trains and the passengers.
These functions are interpreted as follows:
\begin{itemize}
	\item \rm $a(t)$: trains' departure rate from their origin station at time $t$.
	\item \rm $a_p(t)$: passengers' arrival rate at the platform of their origin station at time $t$.
	\item \rm $d(t)$: trains' arrival rate at their final destination station at time $t$.
	\item \rm $d_p(t)$: passengers' arrival rate at their destination station at time $t$.
	\item \rm $T(t)$: travel time of a train and passengers from origin (departs at time $t$) to destination.
		Note that the arrival time at the destination is $t+T(t)$.
\end{itemize}
Therefore, in reality, $a(\cdot)$ and $a_p(\cdot)$ will be determined by the transit operation plan and passenger departure time choice, respectively.
Then, $d(\cdot)$, $d_p(\cdot)$, and $T(\cdot)$ are endogenously determined through the operational dynamics.

In accordance with exit-flow modeling, the train traffic is modeled as follows.
First, the exit flow $d(t)$ is assumed to be
\begin{align}
	d(t) = Q(k(t), a_p(t))	\label{eq_d}
\end{align}
where the FD function $Q(\cdot)$ is considered to be an exit-flow function.\footnotemark{}
\footnotetext{
	If $n_p$ is considered as the sum of the number of passengers who are boarding and alighting (as mentioned in note \ref{note_paxflow}), we can simply define $d(t)$ to be equal to $Q(k(t), a_p(t)+d_p(t))$.
	Such a model is also computable using a similar procedure.
}%
This means that the dynamics of the transit system are modeled by taking the conservation of trains into account as follows:
\begin{align}
	L\frac{\dif k(t)}{\dif t} = a(t) - Q(k(t), a_p(t)),	\label{macrodynamics}
\end{align}
where $L$ represents the length of the transit route.
This exit-flow model has been employed in several studies to represent the macroscopic behavior of a transportation system \cite[e.g.,][]{Merchant1978dta, carey2004exitflow, Daganzo2007gridlock}.
Note that the average train-density $k(t)$ is defined as
\begin{align}
	k(t) = \frac{A(t)-D(t)}{L},	\label{eq_k}
\end{align}
which is consistent with Eq.~\eqref{macrodynamics}.
Based on above functions and equations, $d(t)$ and $D(t)$ are sequentially computed---in other words, the train traffic is computed using the initial and boundary conditions and the exit-flow model based on the FD.

The passenger traffic is derived as follows.
By the definition of the travel time of trains,
\begin{align}
	A(t) = D(t+T(t))	\label{eq_TT}
\end{align}
holds.
As $A(t)$ and $D(t)$ have already been obtained, the travel time $T(t)$ such that Eq.~\eqref{eq_TT} holds is computed.
Then, $D_p(t)$ and $d_p(t)$ are computed from the definition of the travel time of passengers, which is also $T(t)$:
\begin{align}
	&A_p(t) = D_p(t+T(t)).	\label{eq_TTp}
\end{align}

\subsection{Discussion}\label{sec_dynamic_disc}

The proposed macroscopic model computes train out-flow $d(t)$ and passenger out-flow $d_p(t)$ based on the FD function $Q(\cdot)$, the initial and boundary conditions $a(t)$, $a_p(t)$, and $T(0)$.
The notable feature of the model is its high tractability and computational efficiency, as it is based on an exit-flow model.
Therefore, we expect the proposed model to be useful for analyzing various management strategies for transit systems (e.g., dynamic pricing during the morning commute).

It is reasonable to expect that the proposed model can accurately approximate the macroscopic behavior of a transit operation with high-frequency operation (i.e., small time-headway) under moderate changes in demand and/or supply.
This is because exit-flow models are reasonably approximate a dynamical system's behavior when the changes in inflow are moderate compared with the relaxation time of the system.
In the next section, the quantitative accuracy of the model is validated through numerical experiments.

\section{Validation of the Macroscopic Model}\label{sec_valid}

In this section, we validate the quantitative accuracy of the macroscopic model by comparing its results with that of the microscopic model (i.e., Eqs.~\eqref{boarding} and \eqref{newellx}).

\subsection{Simulation Setting}\label{sec_setting}

The parameter values of the transit operation are listed in Table \ref{fdparameters} for both the microscopic and macroscopic models.
The railroad is considered to be a one-way corridor.
The stations are equally spaced at intervals of $l$, and there are a total of 10 stations.
Trains enter the railroad with flow $a(t)$; in the microscopic model, a discrete train enters the railroad from the upstream boundary station if $\lfloor A(t) \rfloor$ (i.e., integer part of $A(t)$) is incremented.
In the microscopic model, trains leave the railroad from the downstream boundary station without any restrictions, other than the passenger boarding and minimum headway clearance.
Passengers arrive at each station with flow $a_p(t)$.

The functions $a(t)$ and $a_p(t)$ are exogenously determined to mimic morning rush hours,
with each having a peak at $t=2$.
The flow before the peak time increases monotonically, whereas the flow after the peak time decreases monotonically---in other words, the so-called S-shaped $A(t)$ and $A_p(t)$ (c.f., \cref{macro_res}) are considered.
The parameters of these functions are the minimum train supply $a^{\min}$, the maximum train supply $a^{\max}$, the minimum passenger demand $a_p^{\min}$, and the maximum passenger demand $a_p^{\max}$.
The functional forms are described in \ref{apndx_sshape}.
The simulation duration is set to 4 h for the baseline scenario in Section \ref{sec_base} and to 8 h for the sensitivity analysis in Section \ref{sec_sens} (the reason will be explained later).

The microscopic model without any control is asymptotically unstable, as proven by \citet{wada2012train_en}; this means that time-varying demand and supply always cause train bunching, making the experiment unrealistic and useless.
Therefore, the headway-based control scheme proposed by \citet{wada2012train_en} is implemented in the microscopic model to prevent bunching and stabilize the operation.
This scheme has two control measures: holding (i.e., extending the dwell time) and an increase of free-flow speed, similar to \citet{Daganzo2009bus}.
The former is activated by a train if its following train is delayed, and is represented as an increase in $g_b$ in the microscopic model.
The latter is activated by a train if it is delayed, and is represented as an increase in $v_f$ up to a maximum allowable speed $v_{\max}$.
In this experiment, $v_{\max}$ is set to 80 km/h and $v_f$ is 70 km/h.
This control scheme can be considered realistic and reasonable, as similar operations are executed in practice.
See \ref{apndx_control} for further details of the control scheme.
Note that the boundary conditions are the trajectory of the first train $x_0(t)~\forall t$, the initial position of all the trains $x_m(0)~\forall m$ (this is converted to the departure time of all the trains from the most upstream station $x_m(t_m^0)~\forall m$ where $t_m^0$ denotes the departure time of train $m$), and the passenger demand to each station $q_p$.

\subsection{Results}\label{sec_res}

First, to examine how well the proposed model reproduces the behavior of the transit system under time-varying conditions, the results for the baseline scenario are presented in Section \ref{sec_base}.
Then, a sensitivity analysis of the demand/supply conditions is conduced and applicable ranges of the proposed model are investigated in Section \ref{sec_sens}.

\subsubsection{Baseline scenario}\label{sec_base}

\begin{figure}[tbp]
	\centering
	\includegraphics[width=0.99\hsize]{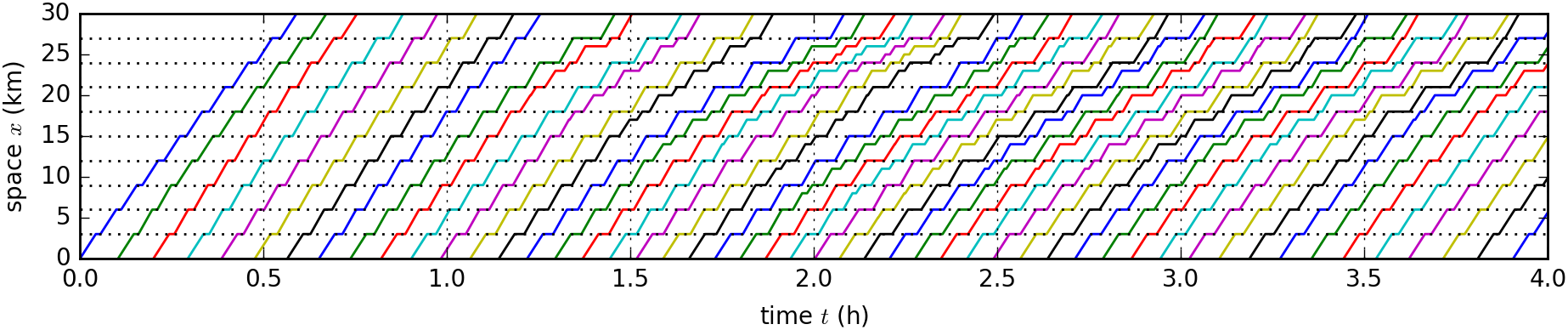}
	\caption{Result of the microscopic model in the baseline scenario.}
	\label{time-space_diagram}
\end{figure}

The baseline scenario with parameter values $a^{\min}=10$ (train/h), $a^{\max}=15$ (train/h), $a_p^{\min}=0.1\mu_p$ (pax/h), and $a_p^{\max}=0.5\mu_p$ (pax/h) is investigated first.
A solution of the microscopic model is shown in Fig.~\ref{time-space_diagram} as a time--space diagram.
The colored curves represent the trajectories of each train traveling in the upward direction while stopping at every station.
Around the peak time period ($t = 2$), train congestion occurs; namely, some of the trains stop occasionally between stations in order to maintain the safety interval.
The congestion is caused by heavy passenger demand; therefore, the situation during rush hour is reproduced.

\begin{figure}[tbp]
	\centering
	\subfloat[Train]{\includegraphics[height=13em]{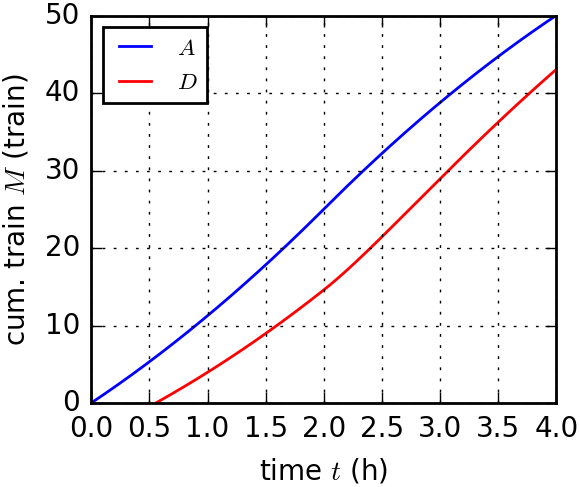}\label{cum_train}}
	\subfloat[Passenger]{\includegraphics[height=13em]{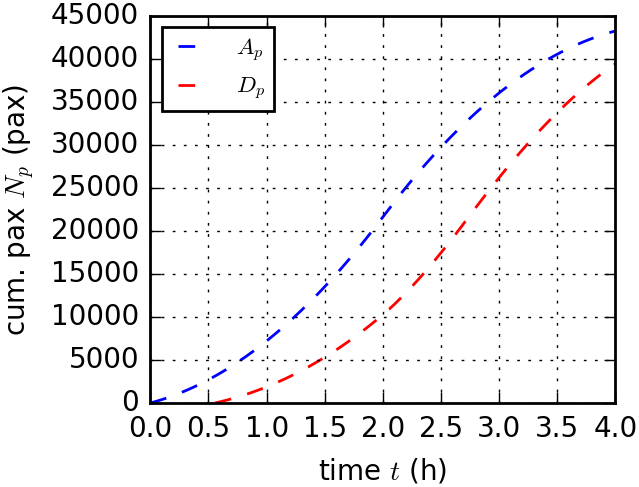}\label{cum_pax}}
	\caption{Result of the macroscopic model in the baseline scenario.}
	\label{macro_res}
\end{figure}

\begin{figure}[tbp]
	\centering
	\includegraphics[height=12em]{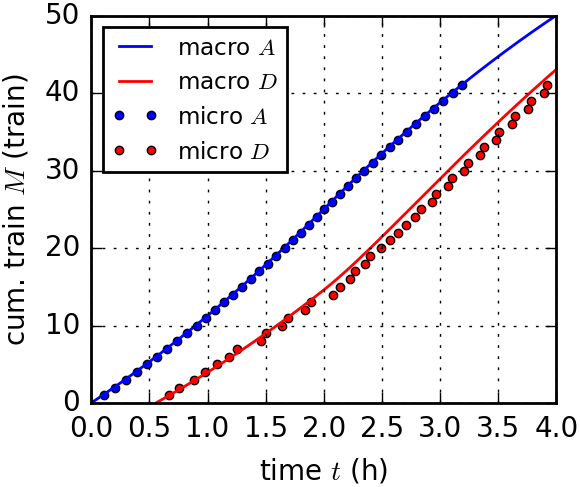}
	\caption{Comparison between the macroscopic and microscopic models in the baseline scenario.}
	\label{compare}
\end{figure}

The result given by the macroscopic model is shown in Fig.~\ref{macro_res} as cumulative plots.
Fig.~\ref{cum_train} shows the cumulative curves for the trains, where the blue curve represents the inflow $A$ and the red curve represents the outflow $D$.
Fig.~\ref{cum_pax} shows those of passengers in the same manner.
Congestion and delay are observed around the peak period (it is more remarkable in the passenger traffic).
For example, during the peak time period, $d_p(t)$ is less than $a_p(t)$ and $a_p(t')$, where $t'$ is time such that $t=t'+T(t')$.
This means that the throughput of the transit system is reduced by the heavy passenger demand.
Consequently, $T(t)$ is greater during peak hours than in off-peak periods such as $T(0)$, meaning that delays occur due to the congestion.

The macroscopic and microscopic models are compared in terms of the cumulative number of trains in Fig.~\ref{compare}.
In the figure, the solid curves denote the macroscopic model and the dots denote the microscopic model.
It is clear that $D$ in the macroscopic model follows that of the microscopic model fairly precisely.
For example, the congestion and delay during the peak time period are captured very well.
However, there is a slight bias: the macroscopic model gives a slightly shorter travel time.
This is mainly due to the large-scale unsteady state (i.e., train bunching) generated in the microscopic model; the delay caused by such large-scale bunching cannot be recovered by the microscopic model under the implemented headway-based control scheme (for details, see \ref{apndx_control}).
It means that if the control is schedule-based, the bias could be reduced.

\subsubsection{Sensitivity analysis of the demand/supply conditions}\label{sec_sens}

The accuracy of the macroscopic model regarding the dynamic patterns of demand/supply is now examined.
This is worth investigating it quantitatively, because it is qualitatively clear that the exit-flow model is valid if the speed of demand/supply changes is ``sufficiently'' small as discussed in Section \ref{sec_dynamic_disc}.
Specifically, the sensitivity of the peak passenger demand $a_p^{\max}$ and train supply $a^{\max}$ is evaluated by assigning various values to these parameters.
The simulation duration is set to 8 h to take the residual delay after $t=4$ (h) in some scenarios into account.
The other parameters are the same as in the baseline scenario.

\begin{figure}[tbp]
	\centering
	\includegraphics[width=.9\hsize]{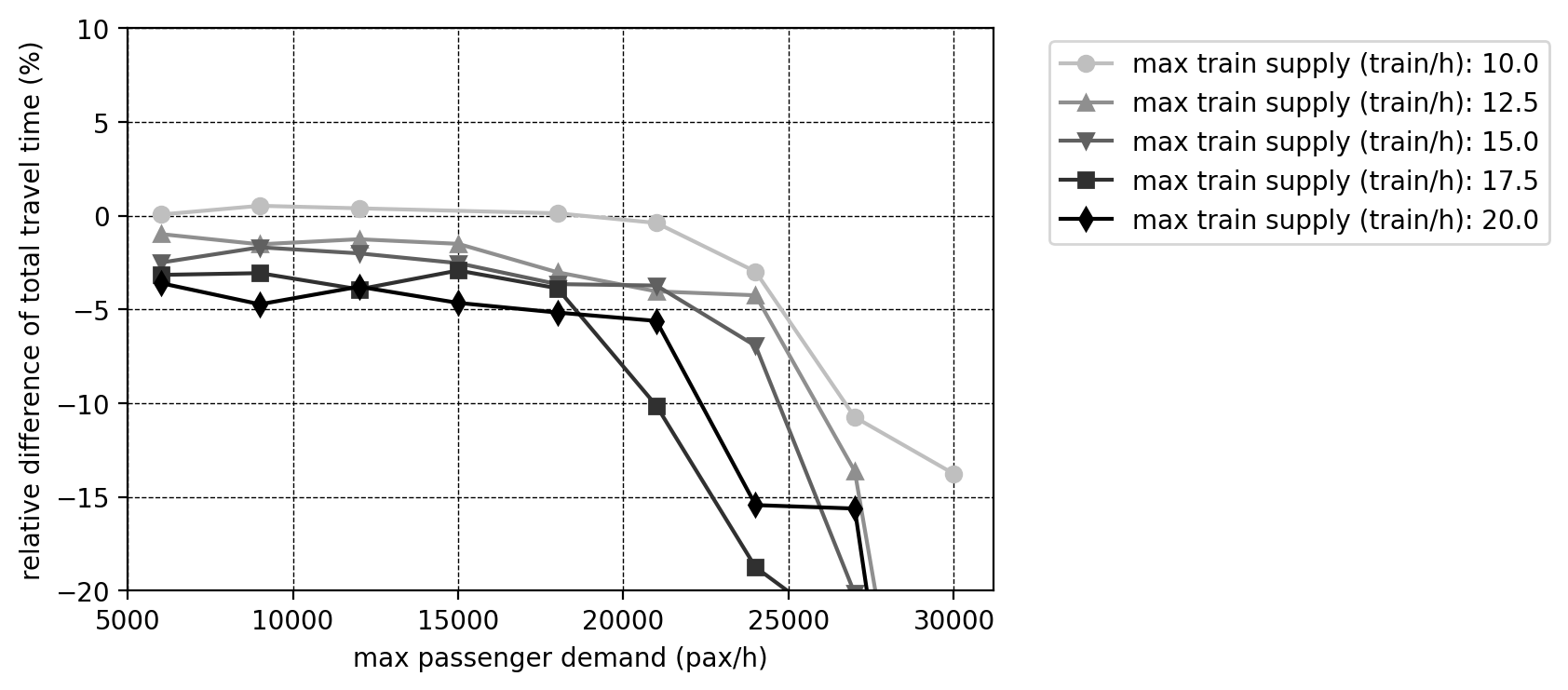}
	\caption{Comparison between the microscopic and macroscopic models under different demand/supply conditions.}
	\label{macro_error}
\end{figure}

The results are summarized in Fig.~\ref{macro_error}.
It shows the relative difference in total travel time (TTT) of trains between the microscopic and macroscopic models for various peak train supply $a^{\max}$ and peak passenger demand $a_p^{\max}$.
The minimum train supply and passenger demand are set as $a^{\min}=10$ (train/h) and $a_p^{\min}=0.1\mu_p = 6000$ (pax/h).
The relative difference can be considered as an error index of the macroscopic model.
The negative values indicate that TTT of the macroscopic model is smaller.

According to the results in Fig.~\ref{macro_error}, the accuracy of the macroscopic model is high when the maximum passenger demand is not extremely large.
This is an expected result, as the speed of demand change is slow in these cases.
TTT given by the macroscopic model is almost always less than that of the microscopic model; this might be due to the aforementioned inconsistency between the steady state assumption of the macroscopic model and headway-based control of the microscopic model.

The relative error increases suddenly when the demand exceeds a certain value, around 20000--22000.
This sudden change is a result of extraordinary large-scale train bunching in the microscopic model.
This bunching often occurs in cases with excessive passenger demand, such as $a_p^{\max} > \mu_p/2$.
Such demand can be considered as unrealistically excessive, as the dwell time of a train at a station is {\it longer} than the cruising time between adjacent stations in such situations; this usually does not occur even in rush hours.

As for the sensitivity of the train supply $a(\cdot)$, there is a weak tendency for faster variations in supply to cause larger errors.
This is also an expected result.
In any case, the error is small.

From these results, we conclude that the proposed model is fairly accurate under ordinary passenger demand, although it is not able to reproduce extraordinary and unrealistic situations with excessive train bunching.
This might be acceptable for representing transit systems during normal rush hours.

\section{Conclusion}\label{sec_conc}

The main contribution of this paper is that it analytically derived a closed-form expression of FD of rail transit systems based on microscopic operation principles.
The FD determines operation performance of rail transit systems (i.e., flow, headway, and mean speed) based on supply of trains and passenger demand.
Furthermore, this paper proposed an efficient, macroscopic dynamic assignment method based on the FD, and numerically showed that the method is fairly accurate under realistic situations.

Specifically, the following three models of an urban rail transit system have been analyzed in this paper:
\begin{itemize}
	\item {\it Microscopic model:} A model describing the trajectories of individual trains and passengers based on Newell's car-following model and passenger boarding model.
		This is represented in Eqs.~\eqref{boarding} and \eqref{newellx}, and is solved using simulations.
	\item {\it Fundamental diagram:} An exact relationship among train-flow, train-density, and passenger-flow in the microscopic model under a steady state.
		This is represented in Eqs.~\eqref{fdeq0}--\eqref{critk}.
		It is a closed-form equation.
	\item {\it Macroscopic model:} A model describing train and passenger traffic using an exit-flow model whose exit-flow function is the FD.
		This is represented in Eqs.~\eqref{eq_d}, \eqref{eq_k}, \eqref{eq_TT}, and \eqref{eq_TTp}, and is solved using simple simulations.
\end{itemize}
The FD and macroscopic model are the original contributions of this study, whereas the microscopic model was proposed by \cite{wada2012train_en}.

The FD itself implies several insights on transit system, such as relation between mean speed of the system and passenger demand.
In addition, according to the results of the numerical experiment, the macroscopic model can reproduce the behavior of the microscopic model accurately, except for cases with unrealistically excessive demands.
Because of the simplicity, mathematical tractability, and good approximation accuracy of the proposed FD and macroscopic model in ordinary situations, it can be expected that they will contribute for obtaining general policy implications on management strategies of rail transit systems, such as pricing and control for morning commute problems.

Following future works are considerable.
First, rigorous empirical validation on the existence of the FD is required.
In fact, several preliminarily results on it have been reported \citep{Fukuda2019trainfd, zhang2019trainfd} as shown in \cref{fd_real}.
Second, as an application of the FD, analysis of operation and demand management for transit systems is important.
For example, the morning commute problem \citep{zhang2021train} has been analyzed, and its departure time choice equilibrium and optimal pricing have been derived.

\appendix

\section{Derivation of FD}\label{apndx_fd}

This appendix describes derivation of the FD expressed in Eqs.~\eqref{fdeq}--\eqref{critk}.
Consider a looped rail transit system under steady state operation.
Let $L$ be the length of the railroad, $S$ be the number of the stations, $M$ be the number of trains, $H$ be the time-headway of the operation, $t_b$ be the dwelling time of a train at a station, $t_c$ be the cruising time of a train between adjacent stations, and $q_p$ be the passenger demand flow rate per station.
Note that the distance between adjacent stations $l$ is $L/S$ and the number of passengers boarding a train at each station is $q_pH$.

The time-headway of the operation is derived as follows.
The round trip time of a train in the looped railroad is $S(t_b+t_c)$, and $M$ trains pass the station during that time.
Then, the identities $NH = S(t_b+t_c)$ and 
\begin{align}
	H = \cfrac{g_{b} + t_c}{M/S - q_{p}/\mu_{p}}	\label{apdfH}
\end{align}
hold.
Moreover, by the definition of headway and Newell's car-following rule, the time-headway $H$ must satisfy
\begin{align}
	H = t_b + \frac{\delta + v\tau}{v} + h_f.	\label{apfdcrit0}
\end{align}
This reduces to
\begin{align}
	H = \cfrac{g_b + \delta/v + \tau}{1-q_p/\mu_p} + h_f.		\label{apfdcrit}
\end{align}

The $q$--$k$ relation in a free-flowing regime is derived as follows.
As the train-flow is $1/H$ and train-density is $M/L$ by definition, Eq.~\eqref{apdfH} is transformed to
\begin{align}
	q(k) = \frac{kl - q_p/\mu_{p}}{g_b + l/v_{f}}.	\label{apfd010}
\end{align}

The train-flow and train-density under a critical state, $(q^*,k^*)$, are derived as follows.
By substituting $v=v_f$ and $h_f=0$ into Eq.~\eqref{apfdcrit} and using the identity $q=k\bar{v}$, we obtain 
\begin{align}
	&q^{*} = \frac{1-q_{p}/\mu_{p}}{g_{b} + \delta/v_{f} + \tau},	\label{apfd0105}\\
	&k^{*}= k_{0} + \frac{(1-q_{p}/\mu_{p})(g_{b} + l/v_{f})}{(g_{b} + \delta/v_{f} + \tau)l},	\label{apfd011}
\end{align}
where $k_0$ is the minimum train-density where the train-flow is zero, namely, $k_{0} = q_{p}/(\mu_{p}l)$.

The $q$--$k$ relation in a congested regime is derived as follows.
First, the $k$--$v$ relation in a congested regime is easily derived from the $q$--$v$ relation \eqref{apfdcrit} with $h_f=0$ and the identity $q=k\bar{v}$:
\begin{align}
	k(v)  = k_{0} + \frac{(1-q_{p}/\mu_{p})(g_{b} + l/v)}{(g_{b} + \delta/v + \tau)l}.
\end{align}
Now, consider $\dif q/ \dif k$, which is identical to $(\dif q/\dif v)\cdot(\dif v/\dif k)$.
This is derived as
\begin{align}
	\cfrac{\dif q}{\dif k} = \cfrac{l\delta}{(\delta - l)g_{b} - \tau l},
\end{align}
which is constant and negative; therefore, the $q$--$k$ relation is linear in a congested regime.
Then, recalling that the linear $q$--$k$ curve passes the point $(q^*,k^*)$ with a slope of $\dif q/ \dif k$, the $q$--$k$ relation in a congested regime is derived as
\begin{align}
 	 q(k) = \cfrac{l\delta}{(\delta - l)g_{b} - \tau l} \ k + q_0	\label{apfd020}
\end{align}
with
\begin{align}
	q_0 = q^{*} - \cfrac{\dif q}{\dif k}\cdot k^{*}.	\label{apfd021}
\end{align}

Eqs.~\eqref{fdeq}--\eqref{critk} are constructed based on Eqs.~\eqref{apfd010}, \eqref{apfd0105}, \eqref{apfd011}, \eqref{apfd020}, and \eqref{apfd021}.

\section{Consistency of the FD and Edie's generalized definition of traffic state}\label{apndx_edie}

It is noteworthy that Eqs.~\eqref{fdeq0} and \eqref{fdeq} are consistent with Edie's generalized definition \citep{Edie1963generalized} of traffic states; because from this consistency we can confirm that the FD is consistent with the fundamental definition of traffic.
For steady-state transit operation, Edie's traffic state is derived as
\begin{align}
	&q = \frac{1}{H},	\label{trainq}\\
	&k = \frac{q_p H/\mu_p+g_b+l/v}{lH},	\label{traink}\\
	&\bar{v} = \frac{l}{q_p H/\mu_p+g_b+l/v}.	\label{trainv}
\end{align}
These relations are derived by applying Edie's definition to the ``minimum component of the time--space diagram'' of the steady state, which is a parallelogram-shaped area in Fig.~\ref{tsd} whose vertexes are time--space points of
(i) train $m$ departs from station $i$,
(ii) train $m$ arrives at station $i+1$,
(iii) train $m-1$ arrives at station $i+1$, and
(iv) train $m-1$ departs from station $i$.
One can easily confirm that Eqs. \eqref{trainq}--\eqref{trainv} satisfy the FD equation.
In fact, the FD equation is also derived from Eqs. \eqref{trainq}--\eqref{trainv} and the constraint \eqref{apfdcrit} induced by Newell's car-following model.

\section{S-shaped supply and demand functions}\label{apndx_sshape}

The train supply and passenger demand in the experiments are given by the following functions:
\begin{align}
	&a(t) = \left\{
		\begin{array}{ll}
			a^{\min}+(a^{\max}-a^{\min})\frac{t}{2},	&	\text{if }	t < 2, \\
			a^{\min}+(a^{\max}-a^{\min})\frac{4-t}{2},	&	\text{if }	2 \leq t < 4,	\\
			a^{\min},	&	\text{if }	4 \leq t,
		\end{array}
	\right.	\\
	&a_p(t) = \left\{
		\begin{array}{ll}
			a_p^{\min}+(a_p^{\max}-a_p^{\min})\frac{t}{2},	&	\text{if }	t < 2, \\
			a_p^{\min}+(a_p^{\max}-a_p^{\min})\frac{4-t}{2}, &	\text{if } 2 \leq t < 4,	\\
			a_p^{\min},	&	\text{if }	4 \leq t.
		\end{array}
	\right.
\end{align}
Both functions have a minimum value at $t=0$ and $t \geq 4$ and a minimum value at $t=2$, and change linearly in between.

\section{Adaptive control scheme in the microscopic model}\label{apndx_control}

This appendix briefly explains the adaptive control scheme for preventing train bunching, proposed by \citet{wada2012train_en}.
This scheme consists of two control measures: holding at a station and increasing the maximum speed during cruising.

First, the scheme modifies the buffer time for dwelling (originally defined as $g_b$ in Eq.~\eqref{boarding}) of train $m$ at station $i$ to
\begin{align}
	g_b := \max\{0,~g_b-E_m(i)\}
\end{align}
with
\begin{align}
	E_m(i) = (1-\alpha)\varepsilon_m(i) + \alpha \mu_p  \left(\varepsilon_m(i) -\varepsilon_{m-1}(i)\right),	\label{wadaEn}
\end{align}
where $\varepsilon_m(i) \equiv t_m(i) - T_{m,i}$ represents the delay, $t_m(i)$ represents the time at which train $m$ arrives at station $i$, $T_{m,i}$ represents the scheduled time (i.e., without delay) at which train $m$ should arrive at station $i$, and $\alpha \in [0,1]$ is a weighting parameter.
This scheme represents a typical holding control strategy, similar to the bunching prevention method of \citet{Daganzo2009bus}, which extends the dwelling time of a vehicle if the headway to the preceding vehicle is too small and vice versa.

Second, the scheme modifies the free-flow cruising speed $v_f$ such that the interstation travel time is reduced by
\begin{align}
	\min\big\{l/v_f-l/v_{\max},~\max\{0,~E_m(i)-g_b\}\big\}.
\end{align}
This means that, in the event of a delay, the train tries to catch up by increasing its cruising speed up to the maximum allowable speed $v_{\max}$ (which implies that the free-flow speed $v_f$ is a ``buffered'' maximum speed).

Meanwhile, the proposed train operation model in this study does not have a schedule---it is a frequency-based operation.
Therefore, in this study, the scheduled headway in the scheme ($T_{m,i}-T_{m-1,i}$) is approximated by the planned frequency ($1/a(t_m(i))$).
Thus, we set $\alpha=1$ and substitute $E_m(i)$ with
\begin{align}
	\mu_p\big(t_m(i)-t_{m-1}(i)-1/a(t_m(i))\big).
\end{align}

The stationary state of the operational dynamics under the original scheme is basically identical to the steady state defined in Section \ref{sec_steady}.
There may be small difference in the congested regime because of the operation scheme; however, this will not be problematic since heavily congested regime will not occur.
In the case of $\alpha < 1$, the scheme makes the train operation asymptotically stable, meaning that the operation schedule is robust to small disturbances.
In the case of $\alpha = 1$, the scheme prevents the propagation and amplification of delay, but does not recover the original schedule (the small `shift' found in Fig.~\ref{compare} is due to $\alpha=1$).
Note that these control measures do not interrupt passenger boarding or violate the safety clearance between trains, meaning that most of the fundamental assumptions of the proposed FD are satisfied.

%

\end{document}